\documentclass[aps,prl,twocolumn,superscriptaddress,10pt]{revtex4-1}

\usepackage{graphicx}
\usepackage{amsmath}
\usepackage{amssymb}
\usepackage{amsfonts}
\usepackage{bm}
\usepackage{psfrag}
\usepackage{pstricks}
\usepackage{color}
\usepackage{soul}
\usepackage{tabularx}
\usepackage{nicefrac}

\pdfoutput=1

\usepackage{hyperref}
\hypersetup{colorlinks=true, linkcolor=blue, citecolor=blue, urlcolor=blue}

\begin{document}

\widetext
\title{Optimized methodology for the calculation of electrostriction from first-principles}


\author{Daniel S.~P.~Tanner}
\email{danielsptanner@gmail.com} 
\affiliation{Universit\'e Paris-Saclay, CentraleSup\'elec, CNRS, Laboratoire SPMS, 91190 Gif-sur-Yvette, France}
\affiliation{Universit\'e de Li\`ege, Q-MAT, CESAM, Institut de Physique}
\author{Eric Bousquet}
\affiliation{Universit\'e de Li\`ege, Q-MAT, CESAM, Institut de Physique}
\author{Pierre-Eymeric Janolin}
\affiliation{Universit\'e Paris-Saclay, CentraleSup\'elec, CNRS, Laboratoire SPMS, 91190 Gif-sur-Yvette, France}
\email{email} 

\date{\today}


\begin{abstract}

In this work we present a new method for the calculation of the electrostrictive properties of materials using density functional theory. The method relies on the thermodynamical equivalence, in a dielectric, of the quadratic mechanical responses (stress or strain) to applied electric stimulus (electric or polarisation fields) to the strain or stress dependence of its dielectric susceptibility or stiffness tensors. Comparing with current finite-field methodologies for the calculation of electrostriction, we demonstrate that our presented methodology offers significant advantages of efficiency, robustness, and ease of use. These advantages render tractable the highthroughput theoretical investigation into the largely unknown electrostrictive properties of materials.

\end{abstract}


\maketitle


Electrostriction is a nonlinear electromechanical coupling present in 
all dielectrics; it is therefore the most ubiquitous electromechanical 
phenomenon.
Despite this, in non-centrosymmetric materials, it is often overshadowed in amplitude  by its linear counterpart: piezoelectricity, which constitutes the primary coupling in most electromechanical systems. 
Electrostrictors are nevertheless employed in sonars~\cite{Pilgrim2016}, actuators~\cite{Uchino1986,Fanson1993} and 
other tunable electromechanical systems.\cite{Newnham1997,Uchino1981,LiJi13} 
There is, in addition, a renewed interest in electrostriction due to the recent observation of electrostrictors that have giant electrostrictive coefficients 10$^4$ 
times larger than the best perovskite materials~\cite{Korobko2012,Li2018,Yuan2018}.

First-principles simulations of electrostriction have not been widely pursued or studied in detail. In this letter, we survey existing methodologies and present a new route to the calculation of electrostrictive properties via density functional theory (DFT), which we show to be more efficient, robust, perspicuous, and easy to apply, than previous methods.

Electrostriction describes the quadratic part of electromechanical couplings that are present in any dielectric material. 
When an electric ($E$) or polarisation ($P$) field is applied to a material, a stress or a strain will be induced, depending on the boundary conditions:
\begin{equation} \label{eq:direct}
  \begin{aligned}
   x_{ij} &=d_{ijk}E_{k} + M_{ijkl}E_{k}E_{l} , \\
   X_{ij} &=e_{ijk}E_{k} + m_{ijkl}E_{k}E_{l} , \\
   x_{ij} &=g_{ijk}P_{k} + Q_{ijkl}P_{k}P_{l} , \\
   X_{ij} &=h_{ijk}P_{k} + q_{ijkl}P_{k}P_{l} . \\
  \end{aligned} 
\end{equation}\\
Here $x_{ij}$ and $X_{ij}$ denote the strain and stress tensor components, respectively; $d_{ijk}$, $e_{ijk}$, $g_{ijk}$, and $h_{ijk}$ the piezoelectric tensors; and $M_{ijkl}$, $m_{ijkl}$, $Q_{ijkl}$, and $q_{ijkl}$  the electrostriction tensors. 
Piezoelectricity is absent in centrosymmetric crystals (as well as in the non-centrosymmetric \textit{432} space group), whereas electrostriction is present in all crystal classes as well as in non-crystalline materials. \cite{NeSu97}

Previous \textit{ab initio} studies of electrostriction\cite{WaMe10,KoWi10,CaFo11,JiZh16,PeKa12, PiKh19,MaHl17,SaRa02} comprise three methodological classes: (i), works which impose a polarisation 
via a frozen polar mode, and subsequently determine the energy-minimising strain, or the energy coupling between strain and polarisation; 
(ii), works which
perform DFT calculations of system properties (such as stress or strain) under the condition of a fixed electric field;\cite{KoWi10,PeKa12} 
and (iii), works which use the recently developed capacity\cite{StSp09} to perform DFT calculations under conditions of fixed displacement field.\cite{CaFo11,JiZh16}

The methods of class (i) utilise various assumptions which limit their applicability to a small class of materials. For example Wang \textit{et al.}~\cite{WaMe10} impose the condition that electrostrictive strains are volume conserving,
which is not true for most materials, including the BaTiO$_{3}$ they study. Of the studies which seek to parametrise Landau-Devonshire potentials\cite{MaHl17,PiKh19}, many do not directly calculate the polarisation using the Berry phase technique, but rather infer it from known values of the Born effective charge tensor or spontaneous polarisations, thus neglecting the electronic contribution to the electrostrictive coefficients. 
Furthermore, fitting to energy differences between different strain states\cite{MaHl17,PiKh19,SaRa02} results in further uncertainties due to a changing basis set,\cite{TaCa19} and increased complexity in fitting equations.

The studies in classes (ii) and (iii) use finite field techniques.\cite{SoIn02,UmPa02,StSp09} 
While these studies do not utilise the same restrictive assumptions 
as those of class (i), there are yet subtleties which must be accounted for.
The first of these is the entanglement of electrostriction and non-linear piezoelectricity in non-centrosymmetric crystals\cite{KoWi10,PeKa12}. 
More generally, all calculations under finite field have a fundamental limitation on k-points and band gap to prevent Zener breakdown.~\cite{SoIn02} 

However, not all viable methods by which to compute electrostriction are represented in the literature.
Here, we aim at presenting and comparing possible methodologies to compute electrostriction from thermodynamic considerations~\cite{Devonshire, NeSu97,LiJi14} and DFT calculations. We exclude methodologies of class (i) mentioned above, as these are not generally applicable. 

First, we present the different possibilities available to compute the electrostrictive coefficients as appearing in Eqs.(\ref{eq:direct}). As mentioned in our review of the literature, and by reference to Eqs.(\ref{eq:direct}), we can see that the electrostrictive coefficients may be obtained by fitting a curve of strain or stress vs electric or polarisation field.\cite{JiZh16,CaFo11,KoWi10,PeKa12}
However, thermodynamic considerations\cite{Devonshire, NeSu97,LiJi14} reveal 
that the four electrostrictive coefficients $M$, $m$, $Q$ and $q$ are related to the partial derivatives of dielectric quantities with respect to mechanical ones as follows:

\begin{equation} \label{eq:converse_Q}
 \begin{aligned}
  &\frac{1}{\epsilon_{0}}\frac{\partial\eta_{ij}}{\partial X_{kl}} = -2Q_{ijkl}   &\frac{1}{\epsilon_{0}}\frac{\partial\eta_{ij}}{\partial x_{kl}} = 2q_{ijkl} \\
  &\epsilon_{0}\frac{\partial\chi_{ij}}{\partial X_{kl}} = 2M_{ijkl} 
  &\epsilon_{0}\frac{\partial\chi_{ij}}{\partial x_{kl}} = -2 m_{ijkl}
  \end{aligned}
\end{equation}

Thus, the coefficients $Q_{ijkl}$ and $q_{ijkl}$ are given by the rate of change of the dielectric stiffness, $\eta_{ij}$ (the inverse of the dielectric susceptibility) with respect to stress, $X_{kl}$, or strain, $x_{kl}$, respectively.
The coefficients $M_{ijkl}$ and $m_{ijkl}$ are given by the rate of change of the susceptibility, $\chi_{ij}$, with respect to stress or strain, respectively.
Density functional perturbation theory (DFPT) enables the calculation of the susceptibility at different stresses or strains, avoiding the problems and disadvantages intrinsic to performing relaxations under a finite $D$ or $E$ field.
Calculations relying on Eqs.~(\ref{eq:converse_Q}) have the following advantages over those relying on 
Eqs.~(\ref{eq:direct}): 
(i) The k-point grid resolution is not limited in a manner dependent on the band gap/field strength. 
(ii) Direct computation of hydrostatic electrostrictive coefficients is possible through the application of hydrostatic strain/pressure. 
This allows one to calculate electrostrictive properties without breaking the intrinsic crystal symmetry, unlike with the imposition of a unidirectional field. 
This means that, for example, 
in centrosymmetric cubic crystals, all four coefficients ($M_{h}$, $m_{h}$, $Q_{h}$, $q_{h}$) can be computed at once without relaxation. 
(iii) The infrastructure to calculate the permitivity/susceptibility using DFPT, which has been available since the 1990s, is more established and robust than that available for energy optimisation in the presence of an applied $E$ or $D$ field
(iv) In piezoelectric materials, the electrostrictive coefficients are still obtained directly from Eqs.~(\ref{eq:converse_Q}), whereas they need to be decoupled from an often much larger piezoelectric effect if one uses Eqs.~(\ref{eq:direct}).\cite{PeKa12}
(v) DFPT allows for the efficient decomposition of the electrostrictive tensors into an electronic and ionic part, and then the decomposition of this ionic part into contributions from each phonon mode.

\begin{figure*}[ht]
   \centering
   \includegraphics[width=\textwidth]{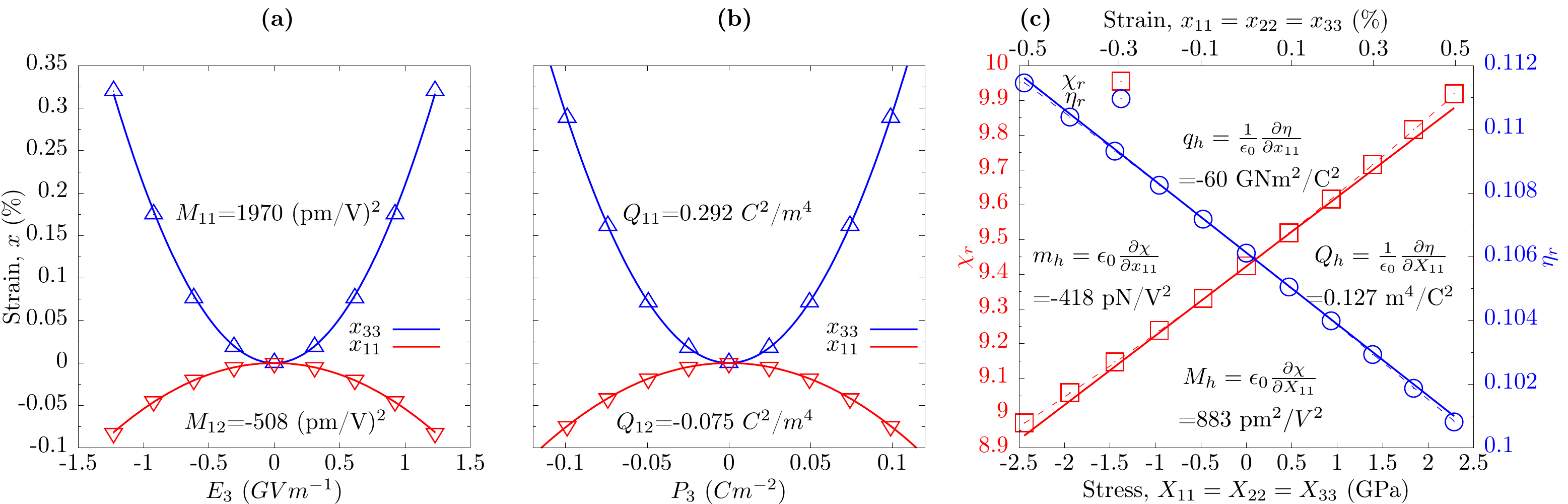}
   \caption{(a) Strain as a function of electric field. (b) Strain as a function of polarisation. Notations have been reduced to two indices: the electric indices have to be identical (i.e. only $M_{ijkk}$ with $k$=1..3); and Voigt notation is used for the mechanical indices. 
   (c) Variation of relative susceptibility $\chi_r$, and relative dielectric stiffness  $\eta_r$ (=$\chi_r^{-1}$) with hydrostatic stress and strain. Calculated $\chi_r$ ($\eta_r$) values are given by red squares (blue circles), with a linear fitting given by the solid red (blue) line, and a quadratic fitting given by a dashed red (blue) line.}
   \label{fig:DirectFittings}
\end{figure*}

To confirm these advantages and validate the method, we have calculated the electrostrictive coefficients using both the methodology proposed here (Eqs.~(\ref{eq:converse_Q})) and the one used so far (Eqs.~(\ref{eq:direct})).
We used the DFT package ABINIT (version 8.6.1) \cite{gonze2016}, with k-point grid densities of 8$\times$8$\times$8 
and a plane wave cutoff energy of 50 Ha, to ensure convergence of electrostrictive coefficients of about 1\%.
We used the PseudoDojo~\cite{Setten2018} normconserving pseudopotentials and the exchange-correlation functional was treated using the generalised gradient approximation of Purdue, Burke, and Ernzerhof, modified for solids, PBEsol.\cite{PeRu08}
We also ensured that the linear and quadratic fittings were appropriate to the given applied strain/electric fields.

In Figure \ref{fig:DirectFittings} we show as an example the results obtained under applied fields for rocksalt MgO.
These were obtained by optimising both the internal atomic positions and lattice vectors at constant $E$ and $D$ fields, for panels (a) and (b), respectively. 
The plots evince the expected quadratic strain versus the $E$ and $P$ fields. 
Electrostriction expands MgO along the direction of the field and contracts it perpendicularly.
Furthermore these plots illustrate that the electrostrictive strains are not volume conserving, contrary to the assumptions of Ref.~\onlinecite{WaMe10} (for example, an electric field of magnitude 1.2~GV/m  will induce a volume expansion of 0.2\%).
The extracted fit value obtained for $M_{11}$ of 1970~pm$^2$/V$^2$ agrees very well with the experimental values of 2020~pm$^2$/V$^2$\cite{SuLi95} (a difference of 2\%). 
We have also calculated the coefficients $q_{11}$, $q_{12}$, $m_{11}$, and $m_{12}$, by fixing the lattice constants, relaxing the internal coordinates under applied $E$ and $D$ fields, and fitting the subsequent stress dependence on $P$ or $E$ field, respectively. These results are summarised  in Table.\ref{tab:CompAndExp} The signs of the $q_{ij}$ and $m_{ij}$ coefficients are opposite to those of the $Q_{ij}$ and $M_{ij}$ coefficients;
this can be expected, a negative axial stress is required to prevent the system from expanding in the direction of the field, and a positive stress
is required to prevent its contraction perpendicular to it. 
To obtain experimental verification for $q_{ij}$ and $m_{ij}$, we use the hydrostatic coefficients $[Q,q,M,m]_{h}=[Q,q,M,m]_{11}+2\times[Q,q,M,m]_{12}$~\cite{NeSu97,LiJi14}, and the relations: $[q,m]_{h}= -3B[Q,M]_{h}$, where $B$ is the bulk modulus. Again, good agreement is found between theory and experiment.

In Fig.~\ref{fig:DirectFittings}~(c) we report the corresponding results obtained using the DFPT calculation of permittivity of MgO under applied hydrostatic strain and stress.
We imposed strains between $\pm 0.5\%$ in steps of 0.1\%. The dielectric susceptibility, stiffness (obtained by inverting the permittivity tensor), and stress at each value of the strain could then be used to determine all four hydrostatic electrostrictive coefficients at once. 
We have also calculated the individual components of the electrostrictive tensors using this method. In this case however, rather than a hydrostatic strain/stress as in Fig.~\ref{fig:DirectFittings}~(c), an axial strain (in Voigt notation) of $x=(\alpha,0,0,0,0,0)$ or axial stress of $X=(\alpha,0,0,0,0,0)$ are imposed on the unit cell, where $\alpha$ varies between $\pm0.3$\% for the strain, and  $\pm2.5$~GPa for the stress. 
We then obtain the electrostriction tensor components by fitting 
with Eqs.~\ref{eq:converse_Q}. 
We found excellent agreement (within 0.1\%) with the hydrostatic coefficients calculated directly from hydrostatic strains/pressure. These results are summarised in Table.\ref{tab:CompAndExp}, where the coefficients subscripted with '$h$' are obtained directly from hydrostatic data, as shown in Fig.~\ref{fig:DirectFittings}~(c).

\begin{table}[h] 
\centering  
\begin{tabularx}{\columnwidth}{>{\setlength\hsize{1\hsize}\centering}X c >{\setlength\hsize{1\hsize}\centering}X c }
\hline\hline
 Coefficient             
 & 
 $\frac{\partial\epsilon_{0}\chi,\nicefrac{\eta}{\epsilon_{0}}}{\partial x,X}$ 
 & 
 $\frac{\partial^{2}x,X}{\partial E^{2},D^{2}}$      
 & 
 Exp       
 \\ 
 \hline
    $m_{11}$                   & -534.7         & -477.7      & -                             \\ 
    $m_{12}$                   & 58.3           & 16.5        & -                             \\ 
    $m_{h}$ (pNV$^{-2}$)       & -418.2         & -444.6      & -396.4\textsuperscript{b,d}    \\ \hline
    $M_{11}$                   & 2251           & 1970        & 2020\textsuperscript{a}                 \\ 
    $M_{12}$                   & -682           & -508        & -                 \\ 
    $M_{h}$ (pm$^{2}$V$^{-2}$) & 883            & 954         & 824\textsuperscript{b}              \\ \hline
    $q_{11}$                    & -76.8          & -71.6      & -                 \\ 
    $q_{12}$                    & 8.4           & 2.5         & -                 \\ 
    $q_{h}$ (GNm$^{2}$C$^{-2}$) & -60.0         & -66.7       & -53.7\textsuperscript{b,d}             \\ \hline
    $Q_{11}$                   & 0.323         & 0.292       & 0.33\textsuperscript{c}                 \\ 
    $Q_{12}$                   & -0.098        & -0.075      & -                 \\ 
    $Q_{h}$ (m$^{4}$C$^{-2}$)  & 0.127         & 0.142       & 0.109\textsuperscript{b}              \\ \hline    
\hline
\end{tabularx}
\caption{Comparison of coefficients obtained via Eqs.~(\ref{eq:direct}) and Eqs.~(\ref{eq:converse_Q}) 
for MgO. 
a=Ref[\onlinecite{SuLi95}]; b=Ref[\onlinecite{BoHa63}];   c=Ref[\onlinecite{Newnham1997}]; d=Ref[\onlinecite{ref1}]. The experimental $M_h$ is that measured in Ref.~[\onlinecite{BoHa63}]; values for $Q_h$, $m_h$, and $q_h$, are obtained from this $M_h$ value using the permittivity of MgO provided by Ref.~[\onlinecite{BoHa63}], and the bulk modulus found in Ref.[\onlinecite{ref1}].} 
\label{tab:CompAndExp}
\end{table} 

The results obtained with our methodology agree with both the ones obtained under applied fields as well as the experimental values as shown in Table \ref{tab:CompAndExp}. Examining the table more closely, we see that Eqs.~(\ref{eq:converse_Q}) produces individual tensor components which are larger in magnitude than those obtained from Eqs.~(\ref{eq:direct}), but that these coefficients sum up to smaller hydrostatic coefficients. The largest disagreement is found for the smaller transverse coefficients, with the disagreement between the two theoretical methods for the hydrostatic coefficients being less than 10\%. Given that previous authors have attributed an uncertainty of 25\% to the finite field method by observing differences in values that should be equal by symmetry\cite{KoWi10}, and that our method exhibits no such differences, 
we attribute the disagreement between the methods to shortcomings of the finite field method. Both methods also show good agreement with experiment, with mean absolute relative errors, over all coefficients, being $<$8\% for each method, which is reasonable agreement, accounting for the large spread found generally in measurements of electrostriction.\cite{ScHa99}

\begin{figure}[htbp]
   \centering
   \includegraphics[width=\columnwidth]{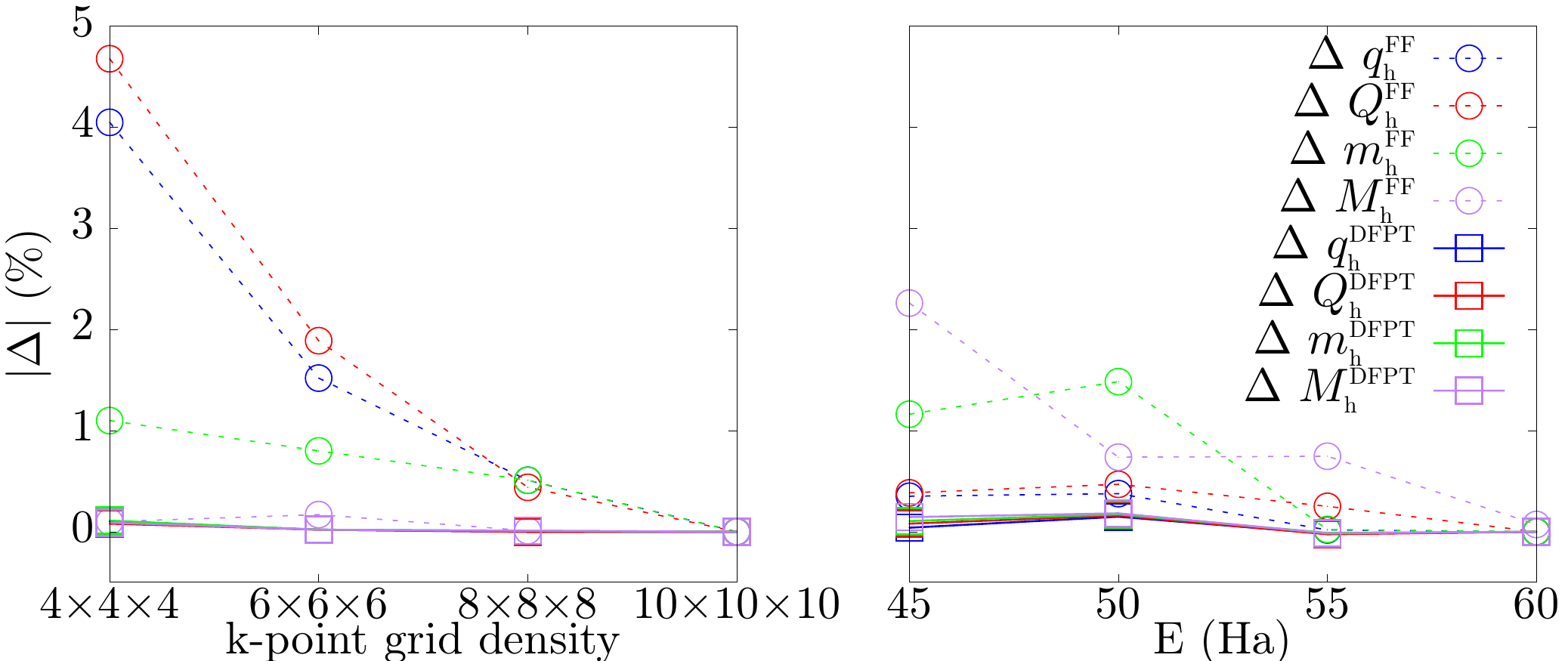}
   \caption{Convergence of calculated electrostrictive properties of rocksalt MgO with respect to (left)
   k-points and (right) planewave cutoff energies.  Data obtained from Eqs.~(\ref{eq:direct}) are given by circles and dashed lines, and data obtained from Eqs.~(\ref{eq:converse_Q}) are given by squares and solid lines. }
   \label{fig:Convergence}
\end{figure}

Having guarantied that the different methods to compute the electrostrictive coefficients give consistent results, we can now compare their efficiencies.
In accordance with its aforementioned advantages, we find that our methodology using DFPT to compute the variations of the permittivity runs significantly faster. 
For example, it is eight times faster to calculate the susceptibility of MgO under a 0.5\% strain than to perform a relaxation under an applied field of $\sim$ 1.25\,GVm$^{-1}$ with the same settings of k-point grid density and cutoff energy.
Furthermore, the calculations using our DFPT methodology also  converge faster with respect to the k-point grid and plane wave cut-off energy as evidenced in Fig.~\ref{fig:Convergence}: 
at a cutoff energy of 45 Ha, all electrostrictive coefficients obtained via our DFPT methodology are within 0.15\% of their values at 60 Ha whereas they are around 2\% with the finite field method.
Likewise with k-point convergence, compared to the  10$\times$10$\times$10 k-point value: about 0.2\% difference with our method versus about 4\% with applied fields with a k-point grid density of 4$\times$4$\times$4.

\begin{figure}[htbp]
   \centering
   \includegraphics[width=0.87\columnwidth]{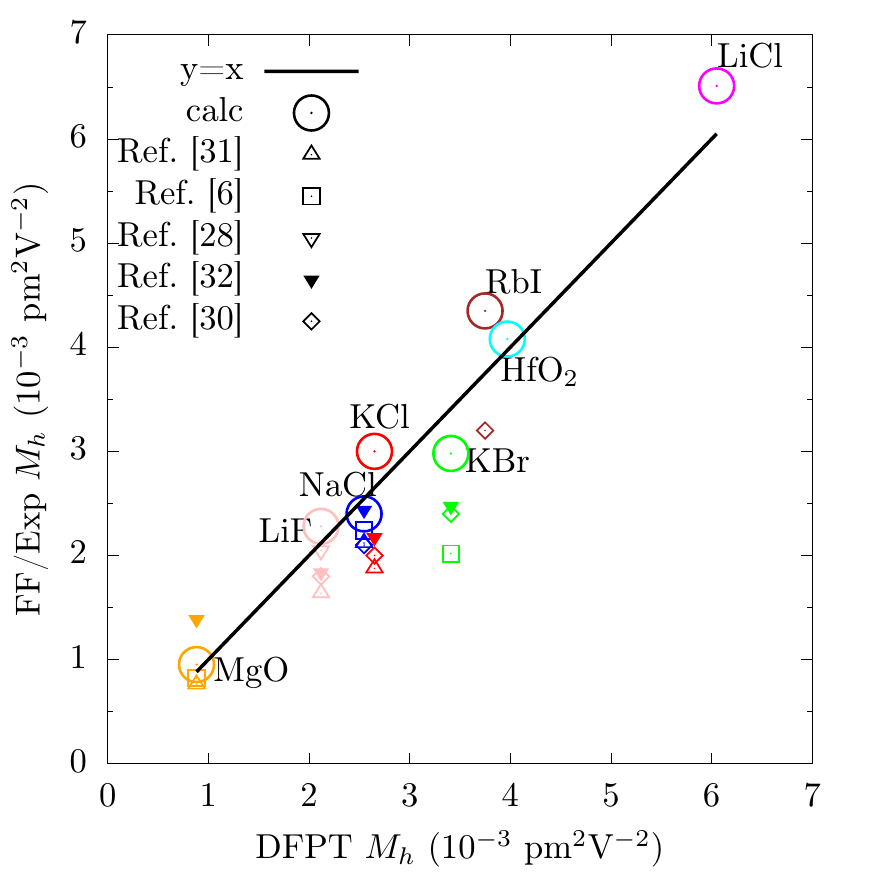}
   \caption{
   Electrostrictive coefficients calculated using Eqs.(\ref{eq:direct}), or measured (denoted by triangles), on the y-axis plotted against calculations based on Eqs.(\ref{eq:converse_Q}) on the x-axis. The line y=x corresponds to exact agreement. Ref~\onlinecite{BaSm73} is low temperature epxeriment based on Eqs.~(\ref{eq:converse_Q}), Refs~\onlinecite{BoHa63,LiJi13, Mayburg1950} are room temperature experiments using Eqs.~(\ref{eq:converse_Q}), and Ref~\onlinecite{ScHa99} is room temperature experiment based on Eqs.~(\ref{eq:direct}). }
   \label{fig:DirectVsConvMh}
\end{figure}

To generalise this validation 
and better corroborate the method, we have calculated electrostrictive properties for a host of materials: rocksalt MgO, LiF, NaCl, KCl, KBr, RbI, and LiCl, and difluorite HfO$_{2}$, using both Eqs.(\ref{eq:direct}) and Eqs.(\ref{eq:converse_Q}). 
In Fig.~\ref{fig:DirectVsConvMh} we plot the calculated $M_h$ electrostrictive coefficients obtained from Eqs.(\ref{eq:direct}) and some experimental values, on the y-axis against the ones from Eqs.(\ref{eq:converse_Q}) on the x-axis. 
Thus, a given material has a fixed x position, and the vertical distance of values from the line y=x represents the extent to which the experimental values, or theoretical finite field obtained values, differ from those calculated using Eqs.(\ref{eq:converse_Q}). We observe almost an order of magnitude spread in the values of the electrostrictive coefficients, LiCl's being the largest.
Given the already large spread in the experimental values, the figure shows excellent agreement between the methods and the general applicability of our methodology.

\begin{figure}[h]
   \centering
   \includegraphics[width=0.9\columnwidth]{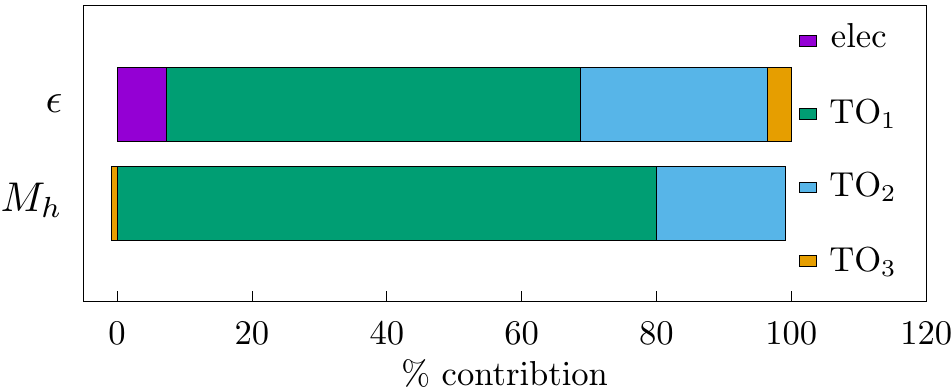}
   \caption{Fractional contributions of electronic and transverse optical phonon degrees of freedom to permittivity $\epsilon$ and electrostrictive tensor $M_{h}$ of BaZrO$_{3}$.}
   \label{fig:Decomposition}
\end{figure}

With the method thus corroborated, we turn now to demonstrate how we may use it to easily obtain a more detailed understanding of electrostriction. 
Indeed, the permittivity tensor can be decomposed into its electronic and ionic contributions, the latter can be further split into each phonon mode contributions. 
The electrostrictive tensors can be decomposed by finding the stress/strain derivatives of these individual components. 
We display such a decomposition for perovskite BaZrO$_{3}$ in Fig.~\ref{fig:Decomposition}, where we show the percentage contribution to the permittivity and electrostrictive coefficient $M_{h}$, of each of their constituents: the electronic part, and the three transverse optical (TO) phonon modes TO$_{1}$ (94 cm$^{-1}$, mode effective charge $\Bar{Z}^*=$ 4.0 e), TO$_{2}$ (190 cm$^{-1}$, $\Bar{Z}^*=$ 5.8 e), and TO$_{3}$ (500 cm$^{-1}$, $\Bar{Z}^*=$ 3.8 e). 
The figure illustrates that the permittivity has a small but non-negligible electronic component of 7.3\%, whilst the largest components are the two first TO phonon modes (61.4\% and 27.8\%, respectively), and the smallest contribution of TO$_3$ mode (3.5\%). 
For the electrostriction, we see that, unlike with the permittivity, the electronic degrees of freedom have a negligible contribution (about 0.01\%, invisible on the plot). 
The two largest contributors are from the TO$_1$ and TO$_2$ modes (80\% and 19\%, respectively), the TO$_{3}$ giving a small but negative contribution (-1\%).
Hence, the largest contribution to the electrostrictive response comes from the softest polar mode with large polarity, in line with the fact that the soft polar mode is extremely sensitive to pressure or strain in ferroelectric perovskites.\cite{KoBe07,JaBo08}
We note that the four coefficients $M_{h}$, $m_{h}$, $Q_{h}$, and $q_{h}$ have the same fractional composition. This follows simply from the linearity of the derivatives with respect to stress or strain, and the equivalence of hydrostatic strain and pressure in a cubic crystal.

In conclusion, we have compared the methodologies for the computation of the electrosctrictive response in crystals using the present common capabilities of DFT codes.
While numerous previous calculations of the electrostrictive coefficients in the literature relied on finite field methods, here we have shown originally that calculating it through the variation of the susceptibility under applied strain (or stress) using DFPT is more convenient, more efficient and faster than the applied fields.
Indeed, our highlighted method not only avoids the drawbacks of applied field methodologies (restriction of k-points density, band gap breakdown) but it appears to be less k-points and plane wave demanding and it requires less computational resources for a given cutoff energy and k-point grid density (by a factor of about 8).
Having validated the method against existing methodologies and experiments, this work thus represents a significant advance in terms of efficiency, robustness, and ease of use, over existing methodologies. This paves the road 
for future high-throughput screening of materials in search of giant electrostrictors, and the investigation of the microscopic origin of giant electrostriction.
A prospective improvement of our methodology would be to have a full DFPT implementation of the electrostrictive response as done for, e.g. the nonlinear optical properties~\cite{veithen2005}.

\section*{Acknowledgements}
Computational resources have been provided by the Consortium des \'Equipements de Calcul Intensif (C\'ECI), funded by the Fonds de la Recherche Scientifique de Belgique (F.R.S.-FNRS) under Grant No. 2.5020.11 and by the Walloon Region.
EB acknowledges FNRS for support and DT aknowledge ULiege Euraxess support.
This work was also performed using HPC resources from the “M\'esocentre” computing centre of CentraleSup\'elec and \'Ecole Normale Sup\'erieure Paris-Saclay supported by CNRS and R\'egion \^Ile-de-France (http://mesocentre.centralesupelec.fr/)
Financial support is acknowledged from a public grant overseen by the French National Research Agency (ANR) as part of the ASTRID program (ANR-19-AST-0024-02). 

\bibliographystyle{apsrev4-1}              
\bibliography     {./Daniel_Tanner_Bibliography} 

\begin{thebibliography}{36}%
\makeatletter
\providecommand \@ifxundefined [1]{%
 \@ifx{#1\undefined}
}%
\providecommand \@ifnum [1]{%
 \ifnum #1\expandafter \@firstoftwo
 \else \expandafter \@secondoftwo
 \fi
}%
\providecommand \@ifx [1]{%
 \ifx #1\expandafter \@firstoftwo
 \else \expandafter \@secondoftwo
 \fi
}%
\providecommand \natexlab [1]{#1}%
\providecommand \enquote  [1]{``#1''}%
\providecommand \bibnamefont  [1]{#1}%
\providecommand \bibfnamefont [1]{#1}%
\providecommand \citenamefont [1]{#1}%
\providecommand \href@noop [0]{\@secondoftwo}%
\providecommand \href [0]{\begingroup \@sanitize@url \@href}%
\providecommand \@href[1]{\@@startlink{#1}\@@href}%
\providecommand \@@href[1]{\endgroup#1\@@endlink}%
\providecommand \@sanitize@url [0]{\catcode `\\12\catcode `\$12\catcode
  `\&12\catcode `\#12\catcode `\^12\catcode `\_12\catcode `\%12\relax}%
\providecommand \@@startlink[1]{}%
\providecommand \@@endlink[0]{}%
\providecommand \url  [0]{\begingroup\@sanitize@url \@url }%
\providecommand \@url [1]{\endgroup\@href {#1}{\urlprefix }}%
\providecommand \urlprefix  [0]{URL }%
\providecommand \Eprint [0]{\href }%
\providecommand \doibase [0]{http://dx.doi.org/}%
\providecommand \selectlanguage [0]{\@gobble}%
\providecommand \bibinfo  [0]{\@secondoftwo}%
\providecommand \bibfield  [0]{\@secondoftwo}%
\providecommand \translation [1]{[#1]}%
\providecommand \BibitemOpen [0]{}%
\providecommand \bibitemStop [0]{}%
\providecommand \bibitemNoStop [0]{.\EOS\space}%
\providecommand \EOS [0]{\spacefactor3000\relax}%
\providecommand \BibitemShut  [1]{\csname bibitem#1\endcsname}%
\let\auto@bib@innerbib\@empty
\bibitem [{\citenamefont {Pilgrim}\ and\ \citenamefont
  {Revathi}(2016)}]{Pilgrim2016}%
  \BibitemOpen
  \bibfield  {author} {\bibinfo {author} {\bibfnamefont {S.}~\bibnamefont
  {Pilgrim}}\ and\ \bibinfo {author} {\bibfnamefont {S.}~\bibnamefont
  {Revathi}},\ }in\ \href {\doibase 10.1016/B978-0-12-803581-8.01716-1} {\emph
  {\bibinfo {booktitle} {Reference Module in Materials Science and Materials
  Engineering}}},\ \bibinfo {series and number} {\bibinfo {number} {August
  2015}}\ (\bibinfo  {publisher} {Elsevier},\ \bibinfo {year} {2016})\ pp.\
  \bibinfo {pages} {1--7}\BibitemShut {NoStop}%
\bibitem [{\citenamefont {Uchino}(1986)}]{Uchino1986}%
  \BibitemOpen
  \bibfield  {author} {\bibinfo {author} {\bibfnamefont {K.}~\bibnamefont
  {Uchino}},\ }\href@noop {} {\bibfield  {journal} {\bibinfo  {journal}
  {American Ceramic Society Bulletin}\ }\textbf {\bibinfo {volume} {65}},\
  \bibinfo {pages} {647} (\bibinfo {year} {1986})}\BibitemShut {NoStop}%
\bibitem [{\citenamefont {Fanson}\ and\ \citenamefont
  {Ealey}(1993)}]{Fanson1993}%
  \BibitemOpen
  \bibfield  {author} {\bibinfo {author} {\bibfnamefont {J.~L.}\ \bibnamefont
  {Fanson}}\ and\ \bibinfo {author} {\bibfnamefont {M.~A.}\ \bibnamefont
  {Ealey}},\ }in\ \href {\doibase 10.1117/12.152675} {\emph {\bibinfo
  {booktitle} {SPIE 1920, Active and Adaptive Optical Components and Systems
  II}}},\ Vol.\ \bibinfo {volume} {1920},\ \bibinfo {editor} {edited by\
  \bibinfo {editor} {\bibfnamefont {M.~A.}\ \bibnamefont {Ealey}}}\ (\bibinfo
  {year} {1993})\ pp.\ \bibinfo {pages} {306--316}\BibitemShut {NoStop}%
\bibitem [{\citenamefont {Newnham}\ \emph
  {et~al.}(1997{\natexlab{a}})\citenamefont {Newnham}, \citenamefont {Sundar},
  \citenamefont {Yimnirun}, \citenamefont {Su},\ and\ \citenamefont
  {Zhang}}]{Newnham1997}%
  \BibitemOpen
  \bibfield  {author} {\bibinfo {author} {\bibfnamefont {R.~E.}\ \bibnamefont
  {Newnham}}, \bibinfo {author} {\bibfnamefont {V.}~\bibnamefont {Sundar}},
  \bibinfo {author} {\bibfnamefont {R.}~\bibnamefont {Yimnirun}}, \bibinfo
  {author} {\bibfnamefont {J.~H.}\ \bibnamefont {Su}}, \ and\ \bibinfo {author}
  {\bibfnamefont {Q.~M.}\ \bibnamefont {Zhang}},\ }\href {\doibase
  10.1021/jp971522c} {\bibfield  {journal} {\bibinfo  {journal} {J. Phys. Chem.
  B}\ }\textbf {\bibinfo {volume} {101}},\ \bibinfo {pages} {10141} (\bibinfo
  {year} {1997}{\natexlab{a}})}\BibitemShut {NoStop}%
\bibitem [{\citenamefont {Uchino}\ \emph {et~al.}(1981)\citenamefont {Uchino},
  \citenamefont {Nomura}, \citenamefont {Cross}, \citenamefont {Newnham},\ and\
  \citenamefont {Jang}}]{Uchino1981}%
  \BibitemOpen
  \bibfield  {author} {\bibinfo {author} {\bibfnamefont {K.}~\bibnamefont
  {Uchino}}, \bibinfo {author} {\bibfnamefont {S.}~\bibnamefont {Nomura}},
  \bibinfo {author} {\bibfnamefont {L.~E.}\ \bibnamefont {Cross}}, \bibinfo
  {author} {\bibfnamefont {R.~E.}\ \bibnamefont {Newnham}}, \ and\ \bibinfo
  {author} {\bibfnamefont {S.~J.}\ \bibnamefont {Jang}},\ }\href {\doibase
  10.1007/BF00552193} {\bibfield  {journal} {\bibinfo  {journal} {Journal of
  Materials Science}\ }\textbf {\bibinfo {volume} {16}},\ \bibinfo {pages}
  {569} (\bibinfo {year} {1981})}\BibitemShut {NoStop}%
\bibitem [{\citenamefont {Li}\ \emph {et~al.}(2014{\natexlab{a}})\citenamefont
  {Li}, \citenamefont {Jin}, \citenamefont {Xu},\ and\ \citenamefont
  {Zhang}}]{LiJi13}%
  \BibitemOpen
  \bibfield  {author} {\bibinfo {author} {\bibfnamefont {F.}~\bibnamefont
  {Li}}, \bibinfo {author} {\bibfnamefont {L.}~\bibnamefont {Jin}}, \bibinfo
  {author} {\bibfnamefont {Z.}~\bibnamefont {Xu}}, \ and\ \bibinfo {author}
  {\bibfnamefont {S.}~\bibnamefont {Zhang}},\ }\href {\doibase
  10.1063/1.4861260} {\bibfield  {journal} {\bibinfo  {journal} {Applied
  Physics Reviews}\ }\textbf {\bibinfo {volume} {1}},\ \bibinfo {pages}
  {011103} (\bibinfo {year} {2014}{\natexlab{a}})}\BibitemShut {NoStop}%
\bibitem [{\citenamefont {Korobko}\ \emph {et~al.}(2012)\citenamefont
  {Korobko}, \citenamefont {Patlolla}, \citenamefont {Kossoy}, \citenamefont
  {Wachtel}, \citenamefont {Tuller}, \citenamefont {Frenkel},\ and\
  \citenamefont {Lubomirsky}}]{Korobko2012}%
  \BibitemOpen
  \bibfield  {author} {\bibinfo {author} {\bibfnamefont {R.}~\bibnamefont
  {Korobko}}, \bibinfo {author} {\bibfnamefont {A.}~\bibnamefont {Patlolla}},
  \bibinfo {author} {\bibfnamefont {A.}~\bibnamefont {Kossoy}}, \bibinfo
  {author} {\bibfnamefont {E.}~\bibnamefont {Wachtel}}, \bibinfo {author}
  {\bibfnamefont {H.~L.}\ \bibnamefont {Tuller}}, \bibinfo {author}
  {\bibfnamefont {A.~I.}\ \bibnamefont {Frenkel}}, \ and\ \bibinfo {author}
  {\bibfnamefont {I.}~\bibnamefont {Lubomirsky}},\ }\href {\doibase
  10.1002/adma.201202270} {\bibfield  {journal} {\bibinfo  {journal} {Advanced
  Materials}\ }\textbf {\bibinfo {volume} {24}},\ \bibinfo {pages} {5857}
  (\bibinfo {year} {2012})}\BibitemShut {NoStop}%
\bibitem [{\citenamefont {Li}\ \emph {et~al.}(2018)\citenamefont {Li},
  \citenamefont {Lu}, \citenamefont {Schiemer}, \citenamefont {Laanait},
  \citenamefont {Balke}, \citenamefont {Zhang}, \citenamefont {Ren},
  \citenamefont {Carpenter}, \citenamefont {Wen}, \citenamefont {Li},
  \citenamefont {Kalinin},\ and\ \citenamefont {Liu}}]{Li2018}%
  \BibitemOpen
  \bibfield  {author} {\bibinfo {author} {\bibfnamefont {Q.}~\bibnamefont
  {Li}}, \bibinfo {author} {\bibfnamefont {T.}~\bibnamefont {Lu}}, \bibinfo
  {author} {\bibfnamefont {J.}~\bibnamefont {Schiemer}}, \bibinfo {author}
  {\bibfnamefont {N.}~\bibnamefont {Laanait}}, \bibinfo {author} {\bibfnamefont
  {N.}~\bibnamefont {Balke}}, \bibinfo {author} {\bibfnamefont
  {Z.}~\bibnamefont {Zhang}}, \bibinfo {author} {\bibfnamefont
  {Y.}~\bibnamefont {Ren}}, \bibinfo {author} {\bibfnamefont {M.~A.}\
  \bibnamefont {Carpenter}}, \bibinfo {author} {\bibfnamefont {H.}~\bibnamefont
  {Wen}}, \bibinfo {author} {\bibfnamefont {J.}~\bibnamefont {Li}}, \bibinfo
  {author} {\bibfnamefont {S.~V.}\ \bibnamefont {Kalinin}}, \ and\ \bibinfo
  {author} {\bibfnamefont {Y.}~\bibnamefont {Liu}},\ }\href {\doibase
  10.1103/PhysRevMaterials.2.041403} {\bibfield  {journal} {\bibinfo  {journal}
  {Physical Review Materials}\ }\textbf {\bibinfo {volume} {2}},\ \bibinfo
  {pages} {041403} (\bibinfo {year} {2018})}\BibitemShut {NoStop}%
\bibitem [{\citenamefont {Yuan}\ \emph {et~al.}(2018)\citenamefont {Yuan},
  \citenamefont {Luna}, \citenamefont {Neri}, \citenamefont {Zakri},
  \citenamefont {Colin},\ and\ \citenamefont {Poulin}}]{Yuan2018}%
  \BibitemOpen
  \bibfield  {author} {\bibinfo {author} {\bibfnamefont {J.}~\bibnamefont
  {Yuan}}, \bibinfo {author} {\bibfnamefont {A.}~\bibnamefont {Luna}}, \bibinfo
  {author} {\bibfnamefont {W.}~\bibnamefont {Neri}}, \bibinfo {author}
  {\bibfnamefont {C.}~\bibnamefont {Zakri}}, \bibinfo {author} {\bibfnamefont
  {A.}~\bibnamefont {Colin}}, \ and\ \bibinfo {author} {\bibfnamefont
  {P.}~\bibnamefont {Poulin}},\ }\href {\doibase 10.1021/acsnano.7b08332}
  {\bibfield  {journal} {\bibinfo  {journal} {ACS Nano}\ }\textbf {\bibinfo
  {volume} {12}},\ \bibinfo {pages} {1688} (\bibinfo {year}
  {2018})}\BibitemShut {NoStop}%
\bibitem [{\citenamefont {Newnham}\ \emph
  {et~al.}(1997{\natexlab{b}})\citenamefont {Newnham}, \citenamefont {Sundar},
  \citenamefont {Yimnirun}, \citenamefont {Su},\ and\ \citenamefont
  {Zhang}}]{NeSu97}%
  \BibitemOpen
  \bibfield  {author} {\bibinfo {author} {\bibfnamefont {R.~E.}\ \bibnamefont
  {Newnham}}, \bibinfo {author} {\bibfnamefont {V.}~\bibnamefont {Sundar}},
  \bibinfo {author} {\bibfnamefont {R.}~\bibnamefont {Yimnirun}}, \bibinfo
  {author} {\bibfnamefont {J.}~\bibnamefont {Su}}, \ and\ \bibinfo {author}
  {\bibfnamefont {Q.~M.}\ \bibnamefont {Zhang}},\ }\href {\doibase
  10.1021/jp971522c} {\bibfield  {journal} {\bibinfo  {journal} {The Journal of
  Physical Chemistry B}\ }\textbf {\bibinfo {volume} {101}},\ \bibinfo {pages}
  {10141} (\bibinfo {year} {1997}{\natexlab{b}})}\BibitemShut {NoStop}%
\bibitem [{\citenamefont {Wang}\ \emph {et~al.}(2010)\citenamefont {Wang},
  \citenamefont {Meng}, \citenamefont {Ma}, \citenamefont {Xu},\ and\
  \citenamefont {Chen}}]{WaMe10}%
  \BibitemOpen
  \bibfield  {author} {\bibinfo {author} {\bibfnamefont {J.~J.}\ \bibnamefont
  {Wang}}, \bibinfo {author} {\bibfnamefont {F.~Y.}\ \bibnamefont {Meng}},
  \bibinfo {author} {\bibfnamefont {X.~Q.}\ \bibnamefont {Ma}}, \bibinfo
  {author} {\bibfnamefont {M.~X.}\ \bibnamefont {Xu}}, \ and\ \bibinfo {author}
  {\bibfnamefont {L.~Q.}\ \bibnamefont {Chen}},\ }\href {\doibase
  10.1063/1.3462441} {\bibfield  {journal} {\bibinfo  {journal} {Journal of
  Applied Physics}\ }\textbf {\bibinfo {volume} {108}},\ \bibinfo {pages}
  {034107} (\bibinfo {year} {2010})}\BibitemShut {NoStop}%
\bibitem [{\citenamefont {Kornev}\ \emph {et~al.}(2010)\citenamefont {Kornev},
  \citenamefont {Willatzen}, \citenamefont {Lassen},\ and\ \citenamefont {Lew
  Yan~Voon}}]{KoWi10}%
  \BibitemOpen
  \bibfield  {author} {\bibinfo {author} {\bibfnamefont {I.}~\bibnamefont
  {Kornev}}, \bibinfo {author} {\bibfnamefont {M.}~\bibnamefont {Willatzen}},
  \bibinfo {author} {\bibfnamefont {B.}~\bibnamefont {Lassen}}, \ and\ \bibinfo
  {author} {\bibfnamefont {L.~C.}\ \bibnamefont {Lew Yan~Voon}},\ }\href
  {\doibase 10.1063/1.3295559} {\bibfield  {journal} {\bibinfo  {journal} {AIP
  Conference Proceedings}\ }\textbf {\bibinfo {volume} {1199}},\ \bibinfo
  {pages} {71} (\bibinfo {year} {2010})}\BibitemShut {NoStop}%
\bibitem [{\citenamefont {Cancellieri}\ \emph {et~al.}(2011)\citenamefont
  {Cancellieri}, \citenamefont {Fontaine}, \citenamefont {Gariglio},
  \citenamefont {Reyren}, \citenamefont {Caviglia}, \citenamefont {F\^ete},
  \citenamefont {Leake}, \citenamefont {Pauli}, \citenamefont {Willmott},
  \citenamefont {Stengel}, \citenamefont {Ghosez},\ and\ \citenamefont
  {Triscone}}]{CaFo11}%
  \BibitemOpen
  \bibfield  {author} {\bibinfo {author} {\bibfnamefont {C.}~\bibnamefont
  {Cancellieri}}, \bibinfo {author} {\bibfnamefont {D.}~\bibnamefont
  {Fontaine}}, \bibinfo {author} {\bibfnamefont {S.}~\bibnamefont {Gariglio}},
  \bibinfo {author} {\bibfnamefont {N.}~\bibnamefont {Reyren}}, \bibinfo
  {author} {\bibfnamefont {A.~D.}\ \bibnamefont {Caviglia}}, \bibinfo {author}
  {\bibfnamefont {A.}~\bibnamefont {F\^ete}}, \bibinfo {author} {\bibfnamefont
  {S.~J.}\ \bibnamefont {Leake}}, \bibinfo {author} {\bibfnamefont {S.~A.}\
  \bibnamefont {Pauli}}, \bibinfo {author} {\bibfnamefont {P.~R.}\ \bibnamefont
  {Willmott}}, \bibinfo {author} {\bibfnamefont {M.}~\bibnamefont {Stengel}},
  \bibinfo {author} {\bibfnamefont {P.}~\bibnamefont {Ghosez}}, \ and\ \bibinfo
  {author} {\bibfnamefont {J.-M.}\ \bibnamefont {Triscone}},\ }\href {\doibase
  10.1103/PhysRevLett.107.056102} {\bibfield  {journal} {\bibinfo  {journal}
  {Phys. Rev. Lett.}\ }\textbf {\bibinfo {volume} {107}},\ \bibinfo {pages}
  {056102} (\bibinfo {year} {2011})}\BibitemShut {NoStop}%
\bibitem [{\citenamefont {Jiang}\ \emph {et~al.}(2016)\citenamefont {Jiang},
  \citenamefont {Zhang}, \citenamefont {Li}, \citenamefont {Jin}, \citenamefont
  {Zhang}, \citenamefont {Wang},\ and\ \citenamefont {Jia}}]{JiZh16}%
  \BibitemOpen
  \bibfield  {author} {\bibinfo {author} {\bibfnamefont {Z.}~\bibnamefont
  {Jiang}}, \bibinfo {author} {\bibfnamefont {R.}~\bibnamefont {Zhang}},
  \bibinfo {author} {\bibfnamefont {F.}~\bibnamefont {Li}}, \bibinfo {author}
  {\bibfnamefont {L.}~\bibnamefont {Jin}}, \bibinfo {author} {\bibfnamefont
  {N.}~\bibnamefont {Zhang}}, \bibinfo {author} {\bibfnamefont
  {D.}~\bibnamefont {Wang}}, \ and\ \bibinfo {author} {\bibfnamefont {C.-L.}\
  \bibnamefont {Jia}},\ }\href {\doibase 10.1063/1.4954886} {\bibfield
  {journal} {\bibinfo  {journal} {AIP Advances}\ }\textbf {\bibinfo {volume}
  {6}},\ \bibinfo {pages} {065122} (\bibinfo {year} {2016})}\BibitemShut
  {NoStop}%
\bibitem [{\citenamefont {Pedesseau}\ \emph {et~al.}(2012)\citenamefont
  {Pedesseau}, \citenamefont {Katan},\ and\ \citenamefont {Even}}]{PeKa12}%
  \BibitemOpen
  \bibfield  {author} {\bibinfo {author} {\bibfnamefont {L.}~\bibnamefont
  {Pedesseau}}, \bibinfo {author} {\bibfnamefont {C.}~\bibnamefont {Katan}}, \
  and\ \bibinfo {author} {\bibfnamefont {J.}~\bibnamefont {Even}},\ }\href
  {\doibase 10.1063/1.3676666} {\bibfield  {journal} {\bibinfo  {journal}
  {Applied Physics Letters}\ }\textbf {\bibinfo {volume} {100}},\ \bibinfo
  {pages} {031903} (\bibinfo {year} {2012})}\BibitemShut {NoStop}%
\bibitem [{\citenamefont {Pitike}\ \emph {et~al.}(2019)\citenamefont {Pitike},
  \citenamefont {Khakpash}, \citenamefont {Mangeri}, \citenamefont {Jr},\ and\
  \citenamefont {Nakhmanson}}]{PiKh19}%
  \BibitemOpen
  \bibfield  {author} {\bibinfo {author} {\bibfnamefont {K.}~\bibnamefont
  {Pitike}}, \bibinfo {author} {\bibfnamefont {N.}~\bibnamefont {Khakpash}},
  \bibinfo {author} {\bibfnamefont {J.}~\bibnamefont {Mangeri}}, \bibinfo
  {author} {\bibfnamefont {G.}~\bibnamefont {Jr}}, \ and\ \bibinfo {author}
  {\bibfnamefont {S.}~\bibnamefont {Nakhmanson}},\ }\href {\doibase
  10.1007/s10853-019-03439-2} {\bibfield  {journal} {\bibinfo  {journal}
  {Journal of Materials Science}\ }\textbf {\bibinfo {volume} {54}},\ \bibinfo
  {pages} {1} (\bibinfo {year} {2019})}\BibitemShut {NoStop}%
\bibitem [{\citenamefont {Marton}\ \emph {et~al.}(2017)\citenamefont {Marton},
  \citenamefont {Kl\'{\i}\ifmmode~\check{c}\else \v{c}\fi{}}, \citenamefont
  {Pa\ifmmode~\acute{s}\else \'{s}\fi{}ciak},\ and\ \citenamefont
  {Hlinka}}]{MaHl17}%
  \BibitemOpen
  \bibfield  {author} {\bibinfo {author} {\bibfnamefont {P.}~\bibnamefont
  {Marton}}, \bibinfo {author} {\bibfnamefont {A.}~\bibnamefont
  {Kl\'{\i}\ifmmode~\check{c}\else \v{c}\fi{}}}, \bibinfo {author}
  {\bibfnamefont {M.}~\bibnamefont {Pa\ifmmode~\acute{s}\else \'{s}\fi{}ciak}},
  \ and\ \bibinfo {author} {\bibfnamefont {J.}~\bibnamefont {Hlinka}},\ }\href
  {\doibase 10.1103/PhysRevB.96.174110} {\bibfield  {journal} {\bibinfo
  {journal} {Phys. Rev. B}\ }\textbf {\bibinfo {volume} {96}},\ \bibinfo
  {pages} {174110} (\bibinfo {year} {2017})}\BibitemShut {NoStop}%
\bibitem [{\citenamefont {Sai}\ \emph {et~al.}(2002)\citenamefont {Sai},
  \citenamefont {Rabe},\ and\ \citenamefont {Vanderbilt}}]{SaRa02}%
  \BibitemOpen
  \bibfield  {author} {\bibinfo {author} {\bibfnamefont {N.}~\bibnamefont
  {Sai}}, \bibinfo {author} {\bibfnamefont {K.~M.}\ \bibnamefont {Rabe}}, \
  and\ \bibinfo {author} {\bibfnamefont {D.}~\bibnamefont {Vanderbilt}},\
  }\href {\doibase 10.1103/PhysRevB.66.104108} {\bibfield  {journal} {\bibinfo
  {journal} {Phys. Rev. B}\ }\textbf {\bibinfo {volume} {66}},\ \bibinfo
  {pages} {104108} (\bibinfo {year} {2002})}\BibitemShut {NoStop}%
\bibitem [{\citenamefont {Stengel}\ \emph {et~al.}(2009)\citenamefont
  {Stengel}, \citenamefont {Spaldin},\ and\ \citenamefont
  {Vanderbilt}}]{StSp09}%
  \BibitemOpen
  \bibfield  {author} {\bibinfo {author} {\bibfnamefont {M.}~\bibnamefont
  {Stengel}}, \bibinfo {author} {\bibfnamefont {N.}~\bibnamefont {Spaldin}}, \
  and\ \bibinfo {author} {\bibfnamefont {D.}~\bibnamefont {Vanderbilt}},\
  }\href {\doibase https://doi.org/10.1038/nphys1185} {\bibfield  {journal}
  {\bibinfo  {journal} {Nature Phys.}\ }\textbf {\bibinfo {volume} {5}},\
  \bibinfo {pages} {304} (\bibinfo {year} {2009})}\BibitemShut {NoStop}%
\bibitem [{\citenamefont {Tanner}\ \emph {et~al.}(2019)\citenamefont {Tanner},
  \citenamefont {Caro}, \citenamefont {Schulz},\ and\ \citenamefont
  {O'Reilly}}]{TaCa19}%
  \BibitemOpen
  \bibfield  {author} {\bibinfo {author} {\bibfnamefont {D.~S.~P.}\
  \bibnamefont {Tanner}}, \bibinfo {author} {\bibfnamefont {M.~A.}\
  \bibnamefont {Caro}}, \bibinfo {author} {\bibfnamefont {S.}~\bibnamefont
  {Schulz}}, \ and\ \bibinfo {author} {\bibfnamefont {E.~P.}\ \bibnamefont
  {O'Reilly}},\ }\href {\doibase 10.1103/PhysRevMaterials.3.013604} {\bibfield
  {journal} {\bibinfo  {journal} {Phys. Rev. Materials}\ }\textbf {\bibinfo
  {volume} {3}},\ \bibinfo {pages} {013604} (\bibinfo {year}
  {2019})}\BibitemShut {NoStop}%
\bibitem [{\citenamefont {Souza}\ \emph {et~al.}(2002)\citenamefont {Souza},
  \citenamefont {\'I\~niguez},\ and\ \citenamefont {Vanderbilt}}]{SoIn02}%
  \BibitemOpen
  \bibfield  {author} {\bibinfo {author} {\bibfnamefont {I.}~\bibnamefont
  {Souza}}, \bibinfo {author} {\bibfnamefont {J.}~\bibnamefont {\'I\~niguez}},
  \ and\ \bibinfo {author} {\bibfnamefont {D.}~\bibnamefont {Vanderbilt}},\
  }\href {\doibase 10.1103/PhysRevLett.89.117602} {\bibfield  {journal}
  {\bibinfo  {journal} {Phys. Rev. Lett.}\ }\textbf {\bibinfo {volume} {89}},\
  \bibinfo {pages} {117602} (\bibinfo {year} {2002})}\BibitemShut {NoStop}%
\bibitem [{\citenamefont {Umari}\ and\ \citenamefont
  {Pasquarello}(2002)}]{UmPa02}%
  \BibitemOpen
  \bibfield  {author} {\bibinfo {author} {\bibfnamefont {P.}~\bibnamefont
  {Umari}}\ and\ \bibinfo {author} {\bibfnamefont {A.}~\bibnamefont
  {Pasquarello}},\ }\href {\doibase 10.1103/PhysRevLett.89.157602} {\bibfield
  {journal} {\bibinfo  {journal} {Phys. Rev. Lett.}\ }\textbf {\bibinfo
  {volume} {89}},\ \bibinfo {pages} {157602} (\bibinfo {year}
  {2002})}\BibitemShut {NoStop}%
\bibitem [{\citenamefont {Devonshire}(1954)}]{Devonshire}%
  \BibitemOpen
  \bibfield  {author} {\bibinfo {author} {\bibfnamefont {A.}~\bibnamefont
  {Devonshire}},\ }\href {\doibase 10.1080/00018735400101173} {\bibfield
  {journal} {\bibinfo  {journal} {Advances in Physics}\ }\textbf {\bibinfo
  {volume} {3}},\ \bibinfo {pages} {85} (\bibinfo {year} {1954})}\BibitemShut
  {NoStop}%
\bibitem [{\citenamefont {Li}\ \emph {et~al.}(2014{\natexlab{b}})\citenamefont
  {Li}, \citenamefont {Jin}, \citenamefont {Xu},\ and\ \citenamefont
  {Zhang}}]{LiJi14}%
  \BibitemOpen
  \bibfield  {author} {\bibinfo {author} {\bibfnamefont {F.}~\bibnamefont
  {Li}}, \bibinfo {author} {\bibfnamefont {L.}~\bibnamefont {Jin}}, \bibinfo
  {author} {\bibfnamefont {Z.}~\bibnamefont {Xu}}, \ and\ \bibinfo {author}
  {\bibfnamefont {S.}~\bibnamefont {Zhang}},\ }\href {\doibase
  10.1063/1.4861260} {\bibfield  {journal} {\bibinfo  {journal} {Applied
  Physics Reviews}\ }\textbf {\bibinfo {volume} {1}},\ \bibinfo {pages}
  {011103} (\bibinfo {year} {2014}{\natexlab{b}})}\BibitemShut {NoStop}%
\bibitem [{\citenamefont {Gonze}\ and\ \citenamefont
  {et~al.}(2016)}]{gonze2016}%
  \BibitemOpen
  \bibfield  {author} {\bibinfo {author} {\bibfnamefont {X.}~\bibnamefont
  {Gonze}}\ and\ \bibinfo {author} {\bibnamefont {et~al.}},\ }\href {\doibase
  https://doi.org/10.1016/j.cpc.2016.04.003} {\bibfield  {journal} {\bibinfo
  {journal} {Computer Physics Communications}\ }\textbf {\bibinfo {volume}
  {205}},\ \bibinfo {pages} {106 } (\bibinfo {year} {2016})}\BibitemShut
  {NoStop}%
\bibitem [{\citenamefont {van Setten}\ \emph {et~al.}(2018)\citenamefont {van
  Setten}, \citenamefont {Giantomassi}, \citenamefont {Bousquet}, \citenamefont
  {Verstraete}, \citenamefont {Hamann}, \citenamefont {Gonze},\ and\
  \citenamefont {Rignanese}}]{Setten2018}%
  \BibitemOpen
  \bibfield  {author} {\bibinfo {author} {\bibfnamefont {M.~J.}\ \bibnamefont
  {van Setten}}, \bibinfo {author} {\bibfnamefont {M.}~\bibnamefont
  {Giantomassi}}, \bibinfo {author} {\bibfnamefont {E.}~\bibnamefont
  {Bousquet}}, \bibinfo {author} {\bibfnamefont {M.~J.}\ \bibnamefont
  {Verstraete}}, \bibinfo {author} {\bibfnamefont {D.~R.}\ \bibnamefont
  {Hamann}}, \bibinfo {author} {\bibfnamefont {X.}~\bibnamefont {Gonze}}, \
  and\ \bibinfo {author} {\bibfnamefont {G.-M.}\ \bibnamefont {Rignanese}},\
  }\href@noop {} {\bibfield  {journal} {\bibinfo  {journal} {Comput. Phys.
  Commun.}\ }\textbf {\bibinfo {volume} {226}},\ \bibinfo {pages} {39}
  (\bibinfo {year} {2018})}\BibitemShut {NoStop}%
\bibitem [{\citenamefont {Perdew}\ \emph {et~al.}(2008)\citenamefont {Perdew},
  \citenamefont {Ruzsinszky}, \citenamefont {Csonka}, \citenamefont {Vydrov},
  \citenamefont {Scuseria}, \citenamefont {Constantin}, \citenamefont {Zhou},\
  and\ \citenamefont {Burke}}]{PeRu08}%
  \BibitemOpen
  \bibfield  {author} {\bibinfo {author} {\bibfnamefont {J.~P.}\ \bibnamefont
  {Perdew}}, \bibinfo {author} {\bibfnamefont {A.}~\bibnamefont {Ruzsinszky}},
  \bibinfo {author} {\bibfnamefont {G.~I.}\ \bibnamefont {Csonka}}, \bibinfo
  {author} {\bibfnamefont {O.~A.}\ \bibnamefont {Vydrov}}, \bibinfo {author}
  {\bibfnamefont {G.~E.}\ \bibnamefont {Scuseria}}, \bibinfo {author}
  {\bibfnamefont {L.~A.}\ \bibnamefont {Constantin}}, \bibinfo {author}
  {\bibfnamefont {X.}~\bibnamefont {Zhou}}, \ and\ \bibinfo {author}
  {\bibfnamefont {K.}~\bibnamefont {Burke}},\ }\href {\doibase
  10.1103/PhysRevLett.100.136406} {\bibfield  {journal} {\bibinfo  {journal}
  {Phys. Rev. Lett.}\ }\textbf {\bibinfo {volume} {100}},\ \bibinfo {pages}
  {136406} (\bibinfo {year} {2008})}\BibitemShut {NoStop}%
\bibitem [{\citenamefont {Sundar}\ \emph {et~al.}(1996)\citenamefont {Sundar},
  \citenamefont {Li}, \citenamefont {Viehland},\ and\ \citenamefont
  {Newnham}}]{SuLi95}%
  \BibitemOpen
  \bibfield  {author} {\bibinfo {author} {\bibfnamefont {V.}~\bibnamefont
  {Sundar}}, \bibinfo {author} {\bibfnamefont {J.-F.}\ \bibnamefont {Li}},
  \bibinfo {author} {\bibfnamefont {D.}~\bibnamefont {Viehland}}, \ and\
  \bibinfo {author} {\bibfnamefont {R.}~\bibnamefont {Newnham}},\ }\href
  {\doibase https://doi.org/10.1016/S0025-5408(96)00036-0} {\bibfield
  {journal} {\bibinfo  {journal} {Materials Research Bulletin}\ }\textbf
  {\bibinfo {volume} {31}},\ \bibinfo {pages} {555 } (\bibinfo {year}
  {1996})}\BibitemShut {NoStop}%
\bibitem [{\citenamefont {Bosman}\ and\ \citenamefont
  {Havinga}(1963)}]{BoHa63}%
  \BibitemOpen
  \bibfield  {author} {\bibinfo {author} {\bibfnamefont {A.~J.}\ \bibnamefont
  {Bosman}}\ and\ \bibinfo {author} {\bibfnamefont {E.~E.}\ \bibnamefont
  {Havinga}},\ }\href {\doibase 10.1103/PhysRev.129.1593} {\bibfield  {journal}
  {\bibinfo  {journal} {Phys. Rev.}\ }\textbf {\bibinfo {volume} {129}},\
  \bibinfo {pages} {1593} (\bibinfo {year} {1963})}\BibitemShut {NoStop}%
\bibitem [{\citenamefont {Madelung}(2004)}]{ref1}%
  \BibitemOpen
  \bibfield  {author} {\bibinfo {author} {\bibfnamefont {O.}~\bibnamefont
  {Madelung}},\ }\href {\doibase 10.1007/978-3-642-18865-7} {\emph {\bibinfo
  {title} {{Semiconductors data handbook; 3rd ed.}}}}\ (\bibinfo  {publisher}
  {Springer},\ \bibinfo {address} {Berlin},\ \bibinfo {year}
  {2004})\BibitemShut {NoStop}%
\bibitem [{\citenamefont {Schreuer}\ and\ \citenamefont
  {Hauss{\"u}hl}(1999)}]{ScHa99}%
  \BibitemOpen
  \bibfield  {author} {\bibinfo {author} {\bibfnamefont {J.}~\bibnamefont
  {Schreuer}}\ and\ \bibinfo {author} {\bibfnamefont {S.}~\bibnamefont
  {Hauss{\"u}hl}},\ }\href@noop {} {\bibfield  {journal} {\bibinfo  {journal}
  {Journal of Physics D}\ }\textbf {\bibinfo {volume} {32}},\ \bibinfo {pages}
  {1263} (\bibinfo {year} {1999})}\BibitemShut {NoStop}%
\bibitem [{\citenamefont {Bartels}\ and\ \citenamefont {Smith}(1973)}]{BaSm73}%
  \BibitemOpen
  \bibfield  {author} {\bibinfo {author} {\bibfnamefont {R.~A.}\ \bibnamefont
  {Bartels}}\ and\ \bibinfo {author} {\bibfnamefont {P.~A.}\ \bibnamefont
  {Smith}},\ }\href {\doibase 10.1103/PhysRevB.7.3885} {\bibfield  {journal}
  {\bibinfo  {journal} {Phys. Rev. B}\ }\textbf {\bibinfo {volume} {7}},\
  \bibinfo {pages} {3885} (\bibinfo {year} {1973})}\BibitemShut {NoStop}%
\bibitem [{\citenamefont {Mayburg}(1950)}]{Mayburg1950}%
  \BibitemOpen
  \bibfield  {author} {\bibinfo {author} {\bibfnamefont {S.}~\bibnamefont
  {Mayburg}},\ }\href {\doibase 10.1103/PhysRev.79.375} {\bibfield  {journal}
  {\bibinfo  {journal} {Phys. Rev.}\ }\textbf {\bibinfo {volume} {79}},\
  \bibinfo {pages} {375} (\bibinfo {year} {1950})}\BibitemShut {NoStop}%
\bibitem [{\citenamefont {Kornev}\ and\ \citenamefont
  {Bellaiche}(2007)}]{KoBe07}%
  \BibitemOpen
  \bibfield  {author} {\bibinfo {author} {\bibfnamefont {I.~A.}\ \bibnamefont
  {Kornev}}\ and\ \bibinfo {author} {\bibfnamefont {L.}~\bibnamefont
  {Bellaiche}},\ }\href {\doibase 10.1080/01411590701228117} {\bibfield
  {journal} {\bibinfo  {journal} {Phase Transitions}\ }\textbf {\bibinfo
  {volume} {80}},\ \bibinfo {pages} {385} (\bibinfo {year} {2007})}\BibitemShut
  {NoStop}%
\bibitem [{\citenamefont {Janolin}\ \emph {et~al.}(2008)\citenamefont
  {Janolin}, \citenamefont {Bouvier}, \citenamefont {Kreisel}, \citenamefont
  {Thomas}, \citenamefont {Kornev}, \citenamefont {Bellaiche}, \citenamefont
  {Crichton}, \citenamefont {Hanfland},\ and\ \citenamefont {Dkhil}}]{JaBo08}%
  \BibitemOpen
  \bibfield  {author} {\bibinfo {author} {\bibfnamefont {P.-E.}\ \bibnamefont
  {Janolin}}, \bibinfo {author} {\bibfnamefont {P.}~\bibnamefont {Bouvier}},
  \bibinfo {author} {\bibfnamefont {J.}~\bibnamefont {Kreisel}}, \bibinfo
  {author} {\bibfnamefont {P.~A.}\ \bibnamefont {Thomas}}, \bibinfo {author}
  {\bibfnamefont {I.~A.}\ \bibnamefont {Kornev}}, \bibinfo {author}
  {\bibfnamefont {L.}~\bibnamefont {Bellaiche}}, \bibinfo {author}
  {\bibfnamefont {W.}~\bibnamefont {Crichton}}, \bibinfo {author}
  {\bibfnamefont {M.}~\bibnamefont {Hanfland}}, \ and\ \bibinfo {author}
  {\bibfnamefont {B.}~\bibnamefont {Dkhil}},\ }\href {\doibase
  10.1103/PhysRevLett.101.237601} {\bibfield  {journal} {\bibinfo  {journal}
  {Phys. Rev. Lett.}\ }\textbf {\bibinfo {volume} {101}},\ \bibinfo {pages}
  {237601} (\bibinfo {year} {2008})}\BibitemShut {NoStop}%
\bibitem [{\citenamefont {Veithen}\ \emph {et~al.}(2005)\citenamefont
  {Veithen}, \citenamefont {Gonze},\ and\ \citenamefont
  {Ghosez}}]{veithen2005}%
  \BibitemOpen
  \bibfield  {author} {\bibinfo {author} {\bibfnamefont {M.}~\bibnamefont
  {Veithen}}, \bibinfo {author} {\bibfnamefont {X.}~\bibnamefont {Gonze}}, \
  and\ \bibinfo {author} {\bibfnamefont {P.}~\bibnamefont {Ghosez}},\ }\href
  {\doibase 10.1103/PhysRevB.71.125107} {\bibfield  {journal} {\bibinfo
  {journal} {Phys. Rev. B}\ }\textbf {\bibinfo {volume} {71}},\ \bibinfo
  {pages} {125107} (\bibinfo {year} {2005})}\BibitemShut {NoStop}%
\end{thebibliography}%

\end{document}